\documentclass[pra]{revtex4}
\usepackage{dcolumn}
\usepackage{bm}
\begin{document}
\title{Ultracold atomic collisions in tight harmonic traps:\\
Quantum defect model and application to metastable helium atoms}
\author{Gillian Peach}
\affiliation{Department of Physics and Astronomy, University College London,
Gower Street, London, WC1E 6BT, UK}
\author{Ian B Whittingham and Timothy J Beams}
\affiliation{School of Mathematical and Physical Sciences, James Cook University,
Townsville, Australia, 4811}
\date{\today}
\begin{abstract}
We analyze a system of two colliding ultracold atoms under strong harmonic
confinement from the viewpoint of quantum defect theory and formulate
a generalized self-consistent method for determining the allowed energies.
We also present two highly efficient computational methods for determining the
bound state energies and eigenfunctions of such systems. The perturbed harmonic
oscillator problem is characterized by a long
asymptotic region beyond the effective range of the interatomic potential.  The
first method, which is based on quantum defect theory and is an adaptation of a
technique developed by one of the authors (GP) for highly excited states in a
modified Coulomb potential, is very efficient for integrating through this outer
region. The second method is a direct numerical solution of the radial
Schr\"odinger equation using a discrete variable representation of the kinetic
energy operator and a scaled radial coordinate grid.  The methods are applied to
the case of trapped spin-polarized metastable helium atoms.
The calculated eigenvalues agree very closely for the two methods,
and with the eigenvalues computed using the generalized self-consistent method.
\end{abstract}

\pacs{03.65.Ge,32.80.Pj,34.50.-s}

\maketitle

\section{Introduction}

An understanding of ultracold collision processes between neutral atoms
is crucial to the design and operation of atom traps, and to the
development of novel quantum processes using trapped atoms \cite{Julienne}.
The elastic collision rate must be high enough to produce efficient
thermalization during the evaporative cooling phase of magnetostatic
trapping, whereas the inelastic collision rate must be small, because such
collisions can generate energetic atoms and change the atomic state, hence
destroying the trapping conditions and producing trap loss.  Elastic
collisions are also important in studies of Bose-Einstein condensates where
they determine the mean field of the condensate \cite{BEC}.  Ultracold
collisions are usually studied under weak trapping conditions in which the
confining inhomogeneous magnetic field is either ignored or is assumed to
have a parabolic or harmonic spatial variation of sufficiently low frequency
(typically $10^{2}$ Hz) that it can be treated as uniform during the
collision.  However, recent interest in phenomena such as quantum phase
transitions of ${}^{87}$Rb atoms confined in three-dimensional optical
lattices \cite{Greiner}, and far off-resonance three-dimensional optical
lattices to store metastable argon atoms \cite{Muller} or to implement
quantum logic gates and create highly entangled quantum states
\cite{qcomputing}, involve conditions where the trapping frequency is
typically $10^{5}$ to $10^{6}$ Hz and the tight trapping environment is
expected to significantly modify the properties of the colliding system.

In several existing calculations for tightly confined neutral atoms
the exact interatomic potential is replaced by the regularised $\delta $-
function pseudopotential
\begin{equation}
\label{vdelta}
V_{\delta}(\bm{r}) = \frac{2 \pi \hbar^{2}}{M} \;a
     \delta (\bm{r}) \frac{\partial}{\partial r} r \, ,
\end{equation}
where $a$ is the scattering length and $M$ is the reduced mass of the system.
This potential reproduces the s-wave phase shifts in the Wigner threshold
regime and also the correct asymptotic behavior of the wavefunction at large
$r$.  This enables an analytical solution for the case of a spherically
symmetric harmonic trap to be obtained \cite{Busch1998}, with the energy
eigenvalues determined from the condition
\begin{equation}
\label{afE}
\frac{a}{\xi} = f(E) \equiv \frac{1}{2} \tan
\left(\frac{\pi E}{2 \hbar \omega} + \frac{\pi}{4}\right)
\frac{\Gamma (\frac{E}{2 \hbar \omega} + \frac{1}{4})}
     {\Gamma (\frac{E}{2 \hbar \omega }+ \frac{3}{4})} \, ,
\end{equation}
where $\omega $ is the trap frequency and $\xi = \sqrt{\hbar /M \omega}$ is
the effective range of the ground state wavefunction.  The validity of this
approach has been investigated by Tiesinga \textit{et al.} \cite{Tiesinga2000}
for the Na and Cs systems by comparing the energy eigenvalues with those
computed numerically using the best available full interatomic potentials.
They find that this approximation is limited to sufficiently weak traps where
$\xi \gg a$.  For the case of two atoms interacting via a hard-sphere potential
of range $a$, Block and Holthaus \cite{Block2002} have shown that the
pseudopotential has the form (\ref{vdelta}) with $a$ given by (\ref{afE}).
Recently, two groups \cite{Bolda2002,Blume2002} have advocated a
model in which an energy-dependent effective scattering length
\begin{equation}
\label{sc}
a_{\text{eff}}(E) = -\frac{\tan \delta_{0}(k)}{k}
\end{equation}
is introduced, where $\hbar k = \sqrt{2 M E}$ and $\delta_0(k)$ is the s-wave
phase shift for the untrapped atoms scattering in the interatomic potential.
The energy eigenvalue condition
\begin{equation}
\label{Bolda}
  \frac{a_{\text{eff}}(E)}{\xi} = f(E)
\end{equation}
is then solved self-consistently.  The procedure given in (\ref{sc}) and
(\ref{Bolda}) reproduces the asymptotic wavefunction and the s-wave phase
shifts even at energies above the Wigner threshold regime, and it is found that
this self-consistent (SC) method works even when $a_{\text{eff}}/\xi \gg 1$.

In this paper we analyze a system of two interacting atoms under strong
harmonic confinement from the viewpoint of quantum defect theory (QDT) and
derive a generalized SC model which validates (\ref{Bolda}) and extends it to
non s-wave collisions.
We then present two highly efficient computational methods for determining the
bound state energies and eigenfunctions for the trapped atoms.  The first
method, based on QDT, is an adaptation of a technique developed by one of the
authors (GP) for highly excited states in a modified Coulomb potential.
The second method is a direct numerical solution of the radial Schr\"odinger
equation using a discrete variable representation (DVR) of the kinetic energy
operator and a scaled radial coordinate grid.  Both methods are applicable to a
system of two arbitrary ultracold neutral atoms interacting in a harmonic trap
and, as an example, we consider the case of spin-polarized metastable helium
tightly confined in a spherically symmetric harmonic trap as this system has
not been studied before under these conditions.

Interest in collision processes in metastable 2$^{3}$S helium (denoted by
He$^{*}$) has been generated by the quest to attain Bose-Einstein condensation
in this system \cite{Santos2001,Robert2001} and, subsequently, to understand
these condensates \cite{Sirjean2002}.  Such condensates are novel in that they
are the first excited-state condensates and open up new fields for
investigation such as those of atomic correlations and the growth kinematics of
the condensate.  This experimental success has depended upon the correctness of
the theoretical prediction
\cite{Shlyap1994,Fedichev1996,Venturi1999,Venturi2000,Leo2001} that the
inelastic Penning ionization processes can be strongly suppressed through
spin polarization of the He$^{*}$ system in the magnetostatic trap.

This paper is organized as follows.  In Sec. II the formalism is developed for
collisions of two neutral atoms in an external three-dimensional isotropic
harmonic trap, and the general nature of the energy eigenvalues and
eigenfunctions of the resultant radial Schr\"{o}dinger equation discussed.
A quantum defect is introduced and shown to be analytic in energy.
In Sec. III we formulate a generalization of the SC method, and the QDT
and DVR computational methods are presented in Sec. IV.  In Sec. V
the QDT, DVR and SC methods are applied to ultracold metastable helium atoms
under tight harmonic confinement.  Results are obtained for s-wave collisions
over a range of trapping frequencies and also for d-wave collisions in a
10 MHz trap.  Finally, in Sec. VI, we summarize and discuss our results.

\section{Two-atom collisions in a harmonic trap}

Consider two atoms $j=1,2$ of mass $M_{j}$ and position $\bm{r}_{j}$ relative
to the centre of the trap.  For a central interatomic potential $V(r)$, where
$r=|\bm{r}|=|\bm{r}_{1}-\bm{r}_{2}|$, and an isotropic harmonic trap, the
two-atom Hamiltonian is separable into centre-of-mass and relative motions.
The equation for the relative motion of angular momentum $l$ is
\begin{equation}
\label{eigen}
\left[ -\frac{\hbar^{2}}{2M} \nabla_{r}^{2}
     + \frac{1}{2}M \omega^{2} r^{2} + V(r)\right] \psi(\bm{r})
     = E \psi(\bm{r}) \, ,
\end{equation}
where $E$ is the energy eigenvalue and the reduced mass
$M =M_{1}M_{2}/(M_{1}+M_{2})$.  The trap potential
$V_{\text{trap}}(r)=M \omega^{2}r^{2}/2$ has been assumed to be independent of
the atomic state, which is generally valid in far-detuned optical lattices
\cite{Grimm}.  As the interaction is spherically symmetric, $\psi(\bm{r})$
has the form
\begin{equation}
\label{psi}
\psi(\bm{r})=\frac{1}{r}F_{kl}(r)Y_{lm}(\theta,\phi) \, ,
\end{equation}
where $Y_{lm}(\theta,\phi)$ are spherical harmonics and $F_{kl}(r)$ satisfies
\begin{equation}
\label{Fkl}
\left[ - \frac{\hbar^{2}}{2 M} \frac{d^{2}}{dr^{2}} +
     \frac{l(l+1)\hbar^{2}}{2 M r^{2}} + \frac{1}{2} M \omega^{2}
     r^{2} + V(r) \right] F_{kl}(r) = E F_{kl}(r) \, .
\end{equation}

It is convenient to introduce the dimensionless variables $\rho = r/\xi $ and
$\kappa = 2E/\hbar \omega =\xi^{2} k^{2}$ and rewrite (\ref{Fkl}) as
\begin{equation}
\label{rho}
\left[ \frac{d^{2}}{d \rho ^{2}}-\frac{l(l+1)}{\rho^{2}}+\kappa
     -\rho^{2}-\frac{2V(\rho)}{\hbar \omega }\right] F(\rho) = 0 \, .
\end{equation}
For large values of $\rho$, $V(\rho)$ is negligible and $F(\rho)$ has
the asymptotic form
\begin{equation}
\label{frho}
      F_a(\rho)=z^{l/2+1/2}\,e^{-z/2}\;w(z) \,;\quad z=\rho^2
\end{equation}
where $w(z)$ is a linearly combination of the two independent solutions
given by
\begin{equation}
\label{w1w2}
     w_{1}(z)={}_{1}F_{1}(a;c;z)\,; \quad
     w_{2}(z)=z^{1-c}{}_{1}F_{1}(1+a-c;2-c;z)
\end{equation}
on using the notation adopted by Luke \cite{Luke}.  In (\ref{w1w2})
\begin{equation}
\label{pars}
\quad a=\frac{l}{2} +\frac{3}{4}-\frac{\kappa}{4}\,;
\quad c=l+\frac{3}{2}
\end{equation}
and ${}_{1}F_{1}(a;c;z)$ is the confluent hypergeometric function
\begin{equation}
\label{confl}
{}_{1}F_{1}(a;c;z)=\sum_{n=0}^{\infty} \frac{\Gamma(a+n)}{\Gamma(a)}\;
\frac{\Gamma(c)}{\Gamma(c+n)}\; \frac{z^{n}}{n!}\, .
\end{equation}
For the case of the unperturbed oscillator the solution $w_{1}(z)$,
which is regular at the origin, is bounded as $z \rightarrow \infty $
provided that $a=-n_{r}$, where $n_{r}$ is a non-negative integer.
The energy eigenvalues are then given by
\begin{equation}
\label{esho0}
      E_0 = \hbar \omega (2 n_{r}+ l + \frac{3}{2})\, ;
\quad n_{r}=0,1,2, \ldots \, ,
\end{equation}
where $n_r$ denotes the number of nodes in the corresponding radial
wavefunction $F_{kl}^{0}(r)$.

In the presence of collisions, the energy eigenvalues $E$ are no longer
equal to $E_0$ but can be written in the form
\begin{equation}
\label{esho}
      E = \hbar \omega (2 n_{r}^{*}+ l + \frac{3}{2})\, .
\end{equation}
Thus the effect of the collisions is to replace $n_{r}$ by
\begin{equation}
\label{defect}
      n_{r}^{*}=n_{r}-\mu
\end{equation}
where $\mu$ will be called a quantum defect and is in general not an
integer; this quantum defect has previously been introduced by
Blume and Greene \cite{Blume2002}.

The present problem, characterized by a long asymptotic trap region
beyond the effective range of the interatomic potential, shows similar
features to the modified Coulomb problem in which a long-range
attractive Coulomb potential is supplemented by a short-range
interaction.  This is an extremely well-known situation in connection
with the analysis of atomic energy levels and spectra where it is
common to specify the energy of a level with quantum numbers $nl$ in
terms of an effective principal quantum number $n^{*}$.  The quantum defect
is then defined to be the difference between $n$ and $n^{*}$.  In this context
there has been extensive development of a quantum defect theory,
see Seaton \cite{Seaton}, and in the following, we demonstrate
that equation (\ref{rho}) for the perturbed harmonic oscillator can be
mathematically transformed into an equation that is identical in form to
that for the modified Coulomb problem.  This enables quantum defect
theory to be directly applied here and valuable theoretical insight into
the underlying behaviour of the energy level structure obtained.

Asymptotically, the exponentially decaying eigenfunction (\ref{frho})
is given by
\begin{equation}
\label{w3}
w_{3}(z)=\psi(a;c;z)=\frac{\Gamma(1-c)}{\Gamma(1+a-c)}\; w_{1}(z)
+ \frac{\Gamma(c-1)}{\Gamma(a)}\; w_{2}(z)
\end{equation}
where $a=-n_r^*$, and as $z \rightarrow \infty $,
\begin{equation}
\label{psiinfty}
\psi(a;c;z) \sim z^{-a} \;{}_{2}F_{0}(a,1+a-c;-\frac{1}{z}) \, ,
\end{equation}
where
\begin{equation}
\label{2f0}
{}_{2}F_{0}(a;c;z) = \sum^{\infty}_{n=0} \frac{\Gamma(a+n)}{\Gamma(a)}\;
\frac{\Gamma(c+n)}{\Gamma(c)}\; \frac{z^{n}}{n!}\, .
\end{equation}

The quantum defect $\mu$ is an analytic function of energy for $E > 0$.  To
prove this we make the transformation
\begin{equation}
\label{farho}
      F_a(\rho)=z^{-1/4}Y(y)
\end{equation}
where $y=\kappa z/8$. The function $Y(y)$ then satisfies
\begin{equation}
\label{coul}
\left[ \frac{d^{2}}{dy^{2}} - \frac{\lambda (\lambda +1)}{y^{2}}
     + \frac{2}{y} + \epsilon \right] Y(y) =0 \, ,
\end{equation}
where $\lambda =l/2-1/4$ and
\begin{equation}
\label{nstar}
\epsilon = -\frac{1}{n^{*2}}\, ;\quad n^{*} \equiv \frac{\kappa}{4} =
     n_{r}^{*} + \lambda + 1 = \nu - \mu
\end{equation}
with $\nu = n_{r}+\lambda +1$.  Equation (\ref{coul}) has exactly the form of
the Coulomb equation for a bound state $nl$, in which $n$ and $l$ have been
replaced by $\nu $ and $\lambda $.  The present problem differs from that
studied by Seaton in that, here, $\lambda $ cannot be an integer.
Consequently we introduce the two linearly independent solutions of
(\ref{coul})
\begin{equation}
\label{y1}
Y_{1}(n^*,\lambda;y) =
\frac{(n^*z)^{\lambda +1}e^{-z/2}}{\Gamma (2 \lambda +2)}
\;{}_{1}F_{1}(\lambda +1-n^{*},2\lambda+2;z)
= \frac{(n^*z)^{\lambda +1}e^{-z/2}}{\Gamma (2 \lambda
+2)}\;w_{1}
\end{equation}
and
\begin{equation}
\label{y2}
Y_{2}(n^*,\lambda;y) = \frac{(n^*z)^{-\lambda}e^{-z/2}}
{\Gamma (-2\lambda )} \;{}_{1}F_{1}(-\lambda -n^{*},-2\lambda;z)
= \frac{(n^{*})^{-\lambda}z^{\lambda +1}e^{-z/2}}
{\Gamma (-2 \lambda )}\;w_{2}\, ,
\end{equation}
where $a = \lambda +1-n^*$ and $c = 2\lambda +2$ in (\ref{w1w2}).  The
functions $Y_{1}$ and $Y_{2}$ are identical to the functions $y_1$ and $y_2$
defined by Seaton and are analytic functions of $\epsilon$.  We then introduce
the general solution
\begin{equation}
\label{y3}
Y_3(n^*,\lambda;y) = z^{\lambda +1}e^{-z/2}
\left[\alpha (n^*)\frac{(n^{*})^{\lambda+1}}{\Gamma (2 \lambda +2)}\;w_{1}
+\beta (n^*)\frac{(n^{*})^{-\lambda}}{\Gamma(-2 \lambda)}\;w_{2}\right] \, ,
\end{equation}
where $\alpha (n^*)$ and $\beta (n^*)$ are analytic functions of $n^*$
($\propto E$), see (\ref{esho}) and (\ref{nstar}).  On expressing $w_{2}$ in
terms of $w_1$ and the exponentially decaying solution $w_{3}$ in (\ref{w3}),
the condition for (\ref{y3}) to decay as $z \rightarrow \infty $ is that the
coefficient of $w_{1}$ must vanish.  This requires that
\begin{equation}
\label{ab}
\alpha (n^*) - A(n^*,\lambda)\,B(n^*,\lambda)\, \beta(n^*) = 0\, ,
\end{equation}
where
\begin{equation}
\label{ab1}
A(n^*,\lambda) =
\frac{\Gamma(\lambda +1+n^*)}{\Gamma(n^*-\lambda)\,(n^{*})^{2\lambda+1}}
\simeq 1 + O(1/n^{*2})
\end{equation}
for large values of $n^*$, and
\begin{equation}
\label{ab2}
B(n^*,\lambda) = \frac{\sin[\pi(n^* + \lambda)]}{\sin[\pi(n^* - \lambda)]}
= (-1)^l \cot{\pi\mu}\, .
\end{equation}
In practice for the cases considered in this paper for which $l\leq 2$ ,
$A(n^*,\lambda)$ remains very close to unity except for the lowest trap states.
Therefore from (\ref{ab})--(\ref{ab2})
\begin{equation}
\alpha(n^*)\sin(\pi\mu) - \beta(n^*)(-1)^l\cos(\pi\mu)\,
[1+O(1/n^{*2})] = 0\, ,
\end{equation}
and so $\mu$ can be written in the form
\begin{equation}
\label{qd1}
\mu = a + bn^* + c n^{*2} + \ldots\;.
\end{equation}

In general, the interatomic potential $V(r)$ supports a number of bound states,
$n_b$ say, for $E < 0$, so that the lowest trap state $(E > 0)$ has $n_r=n_b$,
that is, the lowest trap state eigenfunction has $n_b$ nodes.
If however a pseudopotential is used such as $V_{\delta}(\bm{r})$ in
(\ref{vdelta}), there are no bound states with $E<0$ and the wave function for
the lowest state with $E>0$ has no nodes.  In this case the number of nodes,
$n_r'$, and the quantum defect, $\mu'$, are defined by
\begin{equation}
\label{qd2}
n_r' = n_r - n_b \,; \quad \mu' = \mu -n_b \, .
\end{equation}
The differences between these two types of potential and the effects of their
use are discussed in detail by Peach \cite{Peach}.

\section{Generalization and Validation of the SC method}

Beyond the effective range of the interatomic potential there is an extended
region, typically $10^2 a_0 \leq r \leq 10^4 a_0$, where the effect of the
trapping potential is extremely small and the asymptotic form of the
wavefunction is essentially that of a free wave.  Therefore in this region,
to an excellent approximation $F_{kl}(r)$ can be written as
\begin{equation}
\label{etal}
F_{kl}(r) = kr[\cos \delta_{l}\;j_{l}(kr) - \sin \delta_{l} \;n_{l}(kr)],
\end{equation}
where $j_l(kr)$ and $n_l(kr)$ are spherical Bessel functions, see \cite{Abram},
and $\delta_l(k)$ is the $l$-wave phase shift.  However in the previous
section it has been shown that the correct asymptotic form of the wave
function is obtained from (\ref{farho}) and (\ref{y3}), so that (\ref{etal})
must correspond to
\begin{equation}
\label{FG12}
F_{kl}(r) = C\;F_a(\rho) = C\;z^{-1/4}\left[ \alpha (n^{*}) Y_{1}
+ \beta (n^{*}) Y_{2}\right]\, ,
\end{equation}
where $C$ is a constant.  In equation (\ref{FG12}), the coefficients
$\alpha (n^{*})$ and $\beta (n^{*})$ satisfy equations (\ref{ab})-(\ref{ab2})
and so this form of the solution does decay exponentially at very large
separations as it must for a bound state.  We can ensure that the two
forms (\ref{etal}) and (\ref{FG12}) are identical throughout the region
by examining their behavior for small values of $r$.
The Bessel functions $j_l(kr)$ and $n_l(kr)$ in (\ref{etal}) take the form
\begin{equation}
\label{jnorigin}
   krj_l(kr) \simeq \frac{2^l\Gamma(l+1)}{\Gamma(2l+2)}\;(kr)^{l+1}\,; \quad
   krn_l(kr) \simeq -\frac{\Gamma(2l+1)}{2^l\Gamma(l+1)}\;(kr)^{-l}\, ,
\end{equation}
whereas from (\ref{w1w2}) and (\ref{confl}), for small values
of $z=r^{2}/\xi^{2}$,
\begin{equation}
\label{G12origin}
z^{-1/4}\;Y_{1} \simeq \frac{(n^{*})^{\lambda+1}z^{\lambda+3/4}}
{\Gamma(2\lambda+2)}\,; \quad
z^{-1/4}\;Y_{2} \simeq \frac{(n^{*})^{-\lambda}z^{-\lambda-1/4}}
{\Gamma (-2\lambda)}\, .
\end{equation}
Since $\lambda =l/2-1/4$, the behavior of $F_{kl}(r)$ as
$r \rightarrow 0$ in (\ref{FG12}) is
\begin{equation}
\label{Fklorigin}
F_{kl}(r) \simeq C
\left[\frac{(n^{*})^{\lambda +1}\alpha (n^*)}
{\xi^{l+1}\Gamma(l+\frac{3}{2})}\;
r^{l+1} + \frac{(n^{*})^{-\lambda}\beta (n^*)\xi^{l}}
{\Gamma(-l+\frac{1}{2})}\;
r^{-l}\right],
\end{equation}
 and then from (\ref{etal}), (\ref{jnorigin}) and (\ref{Fklorigin}) we obtain
\begin{equation}
\label{baetal}
\frac{1}{(n^{*})^{2\lambda +1}}\,
\frac{\beta (n^*)}{\alpha (n^*)} = \frac{\tan \delta_l}{(\xi k)^{2l+1}}\;
\frac{\Gamma(-l+\frac{1}{2})}{\Gamma(l+\frac{3}{2})}\;
\frac{\Gamma (2l+1)\Gamma(2l+2)}{[2^{l}\Gamma(l+1)]^{2}}\, 
\end{equation}
where $\xi^2 k^2 = 4 n^*$.  Combining this with the expression for $\beta
(n^*)/\alpha (n^*)$ derived from equations (\ref{ab})-(\ref{ab2}), our final
result is
\begin{equation}
\label{scl}
\frac{\tan \delta_{l}(k)}{(\xi k)^{2l+1}} = - f_{l}(E)\, ,
\end{equation}
where $n^* = E/2\hbar \omega$ and
\begin{eqnarray}
\label{flE}
     f_{l}(E)  =  \frac{1}{2^{2l+1}}
     \tan [\pi(n^* - \frac{l}{2} + \frac{1}{4})]\;
     \frac{\Gamma(n^* - \frac{l}{2} + \frac{1}{4})}
     {\Gamma(n^* + \frac{l}{2} + \frac{3}{4})} \, .
\end{eqnarray}
For $l=0$, (\ref{flE}) reduces to the result given by equations (\ref{sc})
and (\ref{Bolda}).  Thus not only have we obtained a generalization
of the SC method to the case of $l \neq 0$, but we have proved mathematically
that relation (\ref{Bolda}) which was originally introduced empirically,
is in fact rigorously true.

The behavior of (\ref{scl}) for $l>0$ warrants some discussion. For potentials
$V(r)$ with the asymptotic form $r^{-n}$ at large $r$, the threshold
behavior at small $k$ of the phaseshifts $\delta_{l}(k)$ is \cite{Geltman}
\begin{equation}
\label{threshold}
\tan \delta_{l}(k) \propto \left\{
        \begin{array}{ll}
        k^{2l+1}\,;  & n>2l+3  \\
        k^{2l+1}\ln k\,;  &  n=2l+3  \\
        k^{n-2}\,;  & n<2l+3
        \end{array}
        \right. \, .
\end{equation}
In our case, $n=6$ and the behavior as $k \rightarrow 0 $ is given by
\begin{equation}
\label{l01}
\frac{\tan \delta_{l}(k)}{k^{2l+1}} \rightarrow -a_{l}; \quad l=0,1
\end{equation}
and
\begin{equation}
\label{l2}
\frac{\tan \delta_{l}(k)}{k^{4}} \rightarrow b_{l}; \quad l \ge 2 \, ,
\end{equation}
where $a_l$ and $b_l$ are constants.

\section{Computational methods for trapped atoms}

The accurate numerical determination of bound state energies and
eigenfunctions for interacting atoms in a trap requires computational
techniques optimized to deal with both the small inner interatomic region and
the large outer trap region.  Two such methods are now presented in some
detail as they will form the basis of subsequent investigations into more
complex processes involving trapped atoms.

The first approach is an adaptation of a method (QDT) developed by one of us
(GP) for highly excited states in a modified Coulomb potential.  We consider
the region $r\geq r_0 (\approx 20 a_0$ to $40 a_0)$ where the asymptotic solution
specified in (\ref{y3}) is only weakly perturbed by the presence of the
interatomic potential $V(\rho)$.  We first of all generate this asymptotic
solution throughout the region $r\geq r_0$ and then calculate a multiplicative
correction factor in order to obtain an accurate solution of (\ref{coul}).
Equation (\ref{coul}) is written as

\begin{equation}
\label{coula}
\left[ \frac{d^2}{dy^2} - g(y) \right] Y(y) =0 \,; \quad
g(y) = \frac{\lambda (\lambda +1)}{y^2} - \frac{2}{y} - \epsilon \, ,
\end{equation}
and the region is divided up into $N$ ranges, $y = y_n, \, n=0,1,2,
\ldots N$, say.  $y_N$ is chosen so that for $y\geq y_N$ the decaying solution
$Y_3(y)$ can be evaluated by using its asymptotic form, see (\ref{psiinfty})
and (\ref{2f0}).  Within each range $y_{n-1} \leq y \leq y_n$ the following
expansions are made
\begin{equation}
\label{expan}
      g(y) = \sum^{\infty}_{m=0} \frac{(y-y_n)^m}{m!}\,g^{(m)}(y_n)\,;
      \quad Y(y) = \sum^{M_0}_{m=0} a_m (y-y_n)^m
\end{equation}
and substituted into (\ref{coula}).  The coefficients $a_m$ can then be
obtained by equating the coefficients of powers of $(y-y_m)$ to zero, and the
values of $Y(y_{n-1})$ and $Y'(y_{n-1})$ obtained from the range $y_{n-1} \leq
y \leq y_n$ provide the input values for the solution in the next range
$y_{n-2} \leq y \leq y_{n-1}$, etc.  The choice of $M_0$
in each interval clearly depends on its length, but it is found that the
accuracy of the solution $Y(y)$ is very insensitive to the precise choice of
$N$ and $M_0$.  One of the $y_n$ is chosen to be at the outer turning point
given by $g(y)=0, y=y_a$ where
\begin{equation}
\label{ya}
      y_a=n^{*2}\left[1 + \sqrt{1-\frac{\lambda(\lambda +1)}{n^{*2}}}\right]
\end{equation}
and $y_0$ and $\rho_a$ are defined by, see (\ref{coul}) and (\ref{nstar}),
\begin{equation}
\label{y0ya}
      y_0 = n^*(\rho_0)^2/2\,; \quad y_a = n^*(\rho_a)^2/2\, ,
\end{equation}
where $\rho_0 \equiv r_0/\xi$ and $r_a \equiv \xi \rho_a$.
For $y \geq y_a$ (\ref{coula}) is solved to obtain $Y_3(y)$ only, but at
$y = y_a$ the solution $Y_4(y)$ is introduced.  We consider the Airy
functions Ai($q$) and Bi($q$) \cite{Abram}, where $y \approx y_a$ and
\begin{equation}
      q = [g'(y_a)]^{1/3}(y-y_a)\, .
\end{equation}
The functions $Y_3(y_a)$ and $Y_3'(y_a)$ are approximately proportional
to Ai(0) and Ai$'$(0), and a second solution $Y_4(y)$ of (\ref{coul}),
which is exponentially increasing as $y \rightarrow \infty$,
is chosen to be proportional to Bi(0) at $y = y_a$.  Hence
\begin{equation}
\label{y34}
       Y_4(y_a) = \sqrt{3}\, Y_3(y_a)\,; \quad
       Y_4'(y_a) = -\sqrt{3}\, Y_3'(y_a)\, ,
\end{equation}
and then the complex function
\begin{equation}
\label{Y(y)}
       Y_5(y) = Y_3(y) - \text{i} Y_4(y)
\end{equation}
is propagated inwards over the range $y_0 \leq y \leq y_a$.

We now consider the evaluation of the correction factor $R(x)$ where
$x = 1/\rho$.  The solution $F_a(\rho)$ of (\ref{rho}) is written as
\begin{equation}
\label{Rxdef}
       F(\rho) = R(x) F_a(\rho)\, ,
\end{equation}
where $F(\rho)$ is given by (\ref{farho}) in which
\begin{equation}
\label{fy3y5}
      Y(y) \equiv Y_5(y) \, , \quad \rho_0 \leq \rho \leq \rho_a \, ;
      \qquad
      Y(y) \equiv Y_3(y) \, , \quad \rho > \rho_a.
\end{equation}
We also define the functions $\phi(\rho)$ and $\phi_a(\rho)$ by
\begin{equation}
\label{phi}
      \phi(\rho) = \frac{1}{F}\;\frac{dF}{d\rho} \, ; \quad
      \phi_a(\rho) = \frac{1}{F_a(\rho)}\;\frac{dF_a(\rho)}{d\rho}\, .
\end{equation}
The function $R(x)$ then satisfies the equation
\begin{equation}
\label{Rx}
      x^4\frac{d^2 R}{dx^2}+ 2x^3\frac{dR}{dx} - 2 \phi_a(r)\,
      x^2\frac{dR}{dx} - \frac{2 V(x)}{\hbar \omega} R(x) = 0\, ,
\end{equation}
which is solved using the boundary conditions
\begin{equation}
 R(0) =1; \quad \frac{dR}{dx}\Big|_{0}=0 \, .
\end{equation}
By making the choice for $F_a(\rho)$ given by (\ref{fy3y5}) we ensure that
$\phi_0(\rho)$ is a slowly varying function for all $\rho > \rho_0$ and
therefore that $R(x)$ is also a slowly varying function of $x$ over the range
$0 \leq x \leq x_{0}$, where $x_0 = 1/\rho_0$.  Since $r_a$ may be of the order
$10^3 a_0$ to $10^4 a_0$, the range $r_0 \leq r \leq r_a$ can be quite large.
The choice of the complex function here is crucial because it behaves like
$\exp(\text{i} \vartheta)$ where $\vartheta$ varies approximately linearly with
$r$ thus guaranteeing that $\phi_0(\rho)$ changes slowly.  Therefore the
differential equation (\ref{Rx}) for $R(x)$ can always be solved very
accurately throughout the outer region $0 \leq x \leq x_0$, using a grid method
containing a maximum of 66 points and finally the solution $F_{kl}(r)$ required
is obtained by taking the real part of the function in (\ref{Rxdef}).  In
the inner region, $0 <r\leq r_0$, equation (\ref{Fkl}) is integrated
numerically outwards using the Numerov algorithm.  By using an iterative
process to pinpoint the precise value of the energy, the values of $\phi(\rho)$
in (\ref{phi}) obtained from the two regions are matched at $\rho = \rho_0$.
The value of $r_0$ is varied within the range $20 a_0 \leq r_0 \leq 40 a_0$ and the
results are shown to be extremely insensitive to its precise choice.

The second approach to the solution of the energy eigenvalue equation
(\ref{eigen}) is to use a scaled discrete variable representation (DVR)
\cite{Tiesinga1998} (usually a Fourier grid) of the kinetic energy operator,
and convert the bound state problem into one involving the diagonalization of a
$N \times N$ matrix where $N$, for this single-channel problem, is equal to the
number of grid points.  Under a general real invertible transformation of the
radial coordinate $\rho$ given by
\begin{equation}
\label{dvr1}
      t = u(\rho)\, ; \quad \rho = u^{-1}(t) \equiv U(t)
\end{equation}
equation (\ref{rho}) becomes
\begin{equation}
\label{dvr2}
    \left [- f^2 \frac{\text{d}^2}{\text{d}t^2} f^2 + P(t) \right ]
    \tilde F(t) = \kappa \tilde F(t)
\end{equation}
where
\begin{equation}
\label{dvr3}
    f(t) \equiv \left [\frac{\text{d}U}{\text{d}t}\right ]^{-\frac{1}{2}} \, ;
    \quad \tilde F(t) \equiv \frac{F[U(t)]}{f(t)}
\end{equation}
and
\begin{equation}
\label{dvr4}
      P(t) \equiv \frac{l(l+1)}{\rho^{2}} + \rho^{2}
      + \frac{2V(\rho)}{\hbar \omega }
      + f^3(t) \frac{\text{d}^2 f}{\text{d}t^2}\, .
\end{equation}
Equation (\ref{dvr2}) is solved with the boundary conditions $\tilde F(t_1) =
0$ and $\tilde F(t_N) = 0$ by constructing a DVR using a set of basis functions
$\{\phi_m(t)\}$ and a finite set of coordinate points $\{t_m\}$ over the
interval $[t_1,t_N]$.  Then (\ref{dvr2}) is converted into $N$ linear equations
specified by
\begin{equation}
\label{dvr5}
    \sum_{j=1}^N \left [ f^2(t_i) T_{ij} f^2(t_j) + P(t_j)
    \delta_{ij} \right ] \tilde F(t_j) = \kappa \tilde F(t_i)
    \, ; \quad i = 1,2, \ldots N \, .
\end{equation}
For the Fourier basis defined by
\begin{equation}
\label{dvr6}
    \phi_m(t) = \sqrt{\frac{2}{N}} \sin \left [\frac{m\pi(t-t_1)}{t_N-t_1}\right ]
                \, ; \quad m = 1,2, \ldots N
\end{equation}
the DVR of the kinetic energy operator is
\begin{eqnarray}
\label{dvr7}
      T_{ij} = \frac{\pi^2}{2(t_N-t_1)^2} \left\{ 
	\begin{array}{ll}
        (-1)^{i-j} \left [ \csc^2(\frac{\pi}{2N}(i-j))
        - \csc^2(\frac{\pi}{2N}(i+j))\right ]\, ;              & i \neq j  \\
	&  \\
       \frac{1}{3} (2 N^2 + 1) - \csc^2 \left (
       \frac{\pi i}{N}\right )\, ; & i = j 
	\end{array} \right. \, .
\end{eqnarray}
The scaling transformation used 
\begin{equation}
\label{dvr8}
   t = u(\rho) = \left(\frac{\rho}{\zeta}\right)^{\frac{1}{p}}\, ; \quad
   \rho = U(t) = \zeta \, t^p
\end{equation}
involves the parameters $\zeta$ and $\rho$.  Equation (\ref{dvr8})
gives $t_1$ and $t_N$ in terms of $\rho_1$ and $\rho_N$ where theoretically the
boundaries are at $\rho_1 = 0$ and $\rho_N = \infty$.  In practice, because of
the extremely high potential barrier near the origin and the nature of the
scaling, $\rho_1$ is chosen so that $\rho_1\xi = 2 a_0$.  The choice of
$\rho_N$ depends on the number of states being investigated as the associated
eigenfunctions must have exponentially decayed sufficiently on the outer
boundary.  Taking $\rho_N$ = 15 allows us to calculate the positions of about
the first 50 trap states accurately.  The scaling parameters $\zeta$ and $p$
are chosen to ensure that a good proportion of the mesh points are inside the
potential well around the point $r = 7 a_0$.  On taking $\zeta = 20 a_0$ and
$p = 10$ about $17\%$ of the scaled mesh points lie between $r_1$ and $\zeta$,
and with $N = 500$, eight to ten digit convergence is obtained for the
eigenvalues.

\section{Application: Ultracold metastable helium atoms}

As an application of the computational methods adopted in this paper, we
consider the case of spin-polarized metastable helium tightly confined in
harmonic traps of various frequencies.  The colliding atoms are in the
${}^5\Sigma^+_g$ molecular state for which we use the potential of St\"{a}rck
and Meyer \cite{Starck94}.  This potential has a scattering length of
$156.777a_{0}$ and supports 15 bound states.  Calculated quantum defects for
the 31 lowest trap states with $l=0$ are shown in Tables 1 and 2 for trapping
frequencies ranging from 1 to 100 MHz.  Also shown are the results obtained
using the SC solution of (\ref{Bolda}).  Very recently Gad\'ea \textit{et al.}
\cite{Gadea} have constructed a new $^5\Sigma_g$ potential that supports the
same number of states but has a scattering length of $291a_0$.  Experiments
\cite{Santos2001}, \cite{Robert2001} suggest that these two scattering lengths
represent lower and upper limits on the exact value.

The QDT results are in excellent agreement with those obtained from the DVR
method, with the absolute differences being $O(10^{-6})$.  Also the agreement
between the results from the QDT and SC approaches is very good, the absolute
differences being only $O(10^{-7})$ at 1 MHz, increasing to $O(10^{-5})$ for
10, 20 and 50 MHz, and $O(10^{-4})$ at 100 MHz.  For the trapping frequencies
considered, the bound states in the ${}^5\Sigma^+_g$ potential are relatively
unaffected by the presence of the harmonic trap.  The number of bound states
with $E<0$ is still $n_b = 15$, but the most loosely bound state does show some
unusual behavior, being shifted downwards at 1 MHz and upwards at 100 MHz.  The
trapping frequencies of 1, 5, 10, 50 and 100 MHz correspond to the ratios
$a/\xi =$ 0.1167, 0.3691, 0.5221, 0.8255 and 1.167 respectively.  The
pseudopotential result (\ref{afE}), based upon the use of an energy-independent
scattering length $a$, breaks down at the higher trapping frequencies.
Allowance for the variation of $a_{\text{eff}}(E)$ with energy $E$ is essential
in order to obtain the correct eigenvalues; $a_{\text{eff}}(E)$ changes sign as
it passes through a divergence at $k=0.01356 a_{0}^{-1}$, producing a rapid
variation in $a_{\text{eff}}(E)/\xi $ around $E/\hbar \omega \approx 165.6/\nu
(\text{MHz})$.

As a test of the generalized self-consistent (GSC) method (\ref{scl}) we have
considered the case of spin-polarized metastable helium atoms in a 10 MHz trap.
We have chosen $l=2$ since for identical atoms, parity considerations exclude
odd values of $l$, and in this case the $^5\Sigma_g$ potential supports 14
bound states and $b_2=2.383887 \times 10^{5}a_{0}^{4}$ in (\ref{l2}).  Scaled
energy eigenvalues $n^{*}$ calculated for the 31 lowest trap states using the
QDT and GSC methods are given in Table 3.  The two sets of results agree to six
significant figures.  The associated quantum defects $\mu'$ for $l=2$ are much
smaller than the $l=0$ quantum defects, increasing from $O(10^{-4})$ for the
lowest states to $O(10^{-2})$ for the highest states considered.

\section{Summary and discussion}

A QDT analysis of colliding ultracold atoms tightly confined in harmonic
potentials has been undertaken and a generalized SC model obtained.
Two highly efficient computational methods are presented for calculating
the bound state energies and eigenfunctions,
one based upon QDT and the other on a scaled DVR method.  The perturbed
harmonic oscillator problem is characterized by a long asymptotic region
beyond the effective range of the interatomic potential and the QDT method is
very efficient for integrating inwards through this outer region.
The radial Schr\"{o}dinger equation for the relative motion of the harmonically
confined atoms is transformed into that for a modified Coulomb potential and a
quantum defect introduced that is an analytic function of energy.  At large
separations $r$, each eigenfunction for the confined atoms is expressed as a
product of the unperturbed harmonic oscillator eigenfunction and a residual
function $R(x)$ that is slowly varying in $x=1/r$.  The unperturbed function is
calculated by dividing the region into a number of ranges and expanding the
function in a power series within each range.  The differential equation for
$R(x)$ is then solved accurately using a grid method requiring relatively few
mesh points.  The Schr\"{o}dinger equation is integrated directly through the
inner region and the two solutions matched at an intermediate separation
in the range 20 to 40\,$a_0$.
The eigenvalues are determined very efficiently by iteration on an initial
estimate obtained by extrapolating the quantum defects downwards in energy
from highly excited trap states.

The two computational methods have been applied to the case of trapped
spin-polarized metastable helium atoms.
Energy eigenvalues calculated with the QDT method agree closely with those
computed directly from the radial Schr\"{o}dinger equation for the trapped
atoms using a DVR method, and with those obtained using a self-consistent
method (SC) involving an energy-dependent effective scattering length.

The range of trapping frequencies and the number of trap states considered in the
present investigation was motivated by the desire to test the validity and robustness of
our computational methods and not by current experimental feasibility. The higher trap
frequencies are beyond those presently used and the higher trap states
may  not be physically realistic.
The harmonic approximation to the confining potential of the optical lattice is
only valid for the lower lying states of the atoms confined near the nodes or
antinodes of the lattice.  Also, the rate of quantum tunnelling from these
higher states to neighboring wells in the lattice may be significant, thus
requiring a calculation of the band structure type for the entire lattice.

The present calculations need to be extended to include collisional loss
processes and to study the loss rate as a function of trapping frequency.  In
particular, for trapped metastable helium atoms, it will be important to
include the magnetic dipole-dipole interactions that couple the spin-polarized
${}^5\Sigma^+_g$ state to the ${}^1\Sigma^+_g$ state from which there is a high
probability of loss through Penning and associative ionization at small
interatomic separations.
 
\newpage
\begingroup
\squeezetable
\begin{table}
\caption{Quantum defects $\mu'=\mu-15$ for the lowest states with $E>0$ and
$l=0$ where $n_r'=n_r-15$.  Results are listed for 1, 10 and 20 MHz harmonic
traps calculated using quantum defect theory (QDT) and the self-consistent
method (SC).}
\begin{ruledtabular}
\begin{tabular}{ldddddd}
$n_r'$   & \multicolumn{2}{c}{1 MHz} & \multicolumn{2}{c}{10 MHz}
& \multicolumn{2}{c}{20 MHz} \\
  &  \multicolumn{1}{c}{QDT}  &  \multicolumn{1}{c}{SC}   &
  \multicolumn{1}{c}{QDT}  &  \multicolumn{1}{c}{SC}  &
  \multicolumn{1}{c}{QDT}  &  \multicolumn{1}{c}{SC}  \\
\hline

0  & -0.06814498  & -0.068145  & -0.2202828  &  -0.220309
   & -0.3057509  & -0.305860
 \\
1  & -0.1002427  & -0.100243  & -0.3009027  &  -0.300927
   & -0.3996139  & -0.399707
 \\
2  & -0.1237965  & -0.123797  & -0.3540177  &  -0.354041
   & -0.4593226  & -0.459905
\\
3  & -0.1430405  & -0.143040  & -0.3941769  &  -0.394199
   & -0.5038833  & -0.503958
\\
4  & -0.1595646  & -0.159564  & -0.4266773  &  -0.426698
   & -0.5398083  & -0.539876
\\
5  & -0.1741690  & -0.174169  & -0.4540938  &  -0.454114
   & -0.5701305  & -0.570193
\\
6  & -0.1873241  & -0.187324  & -0.4778824  &  -0.477901
   & -0.5965138  & -0.596557
\\
7  & -0.1993344  & -0.199334  & -0.4989491  &  -0.498965
   & -0.6199703  & -0.620024
\\
8  & -0.2104109  & -0.210411  & -0.5178967  &  -0.517914
   & -0.6411619  & -0.641230
\\
9  & -0.2207071  & -0.220707  & -0.5351471  &  -0.535164
   & -0.6605457  & -0.660645
\\
10 & -0.2303387  & -0.230339  & -0.5510070  &  -0.551023
   & -0.6784511  & -0.678534
\\
11 & -0.2393959  & -0.239396  & -0.5657064  &  -0.565722
   & -0.6951230  & -0.695193
\\
12 & -0.2479504  & -0.247951  & -0.5794221  &  -0.579437
   & -0.7107489  & -0.710810
\\
13 & -0.2560604  & -0.256060  & -0.5922930  &  -0.592307
   & -0.7254758  & -0.725529
\\
14 & -0.2637738  & -0.263774  & -0.6044302  &  -0.604444
   & -0.7394208  & -0.739466
\\
15 & -0.2711310  & -0.271132  & -0.6159241  &  -0.615936
   & -0.7526786  & -0.752716
\\
16 & -0.2781659  & -0.278166  & -0.6268491  &  -0.626862
   & -0.7653274  & -0.765366
\\
17 & -0.2849077  & -0.284908  & -0.6372671  &  -0.637285
   & -0.7774321  & -0.777473
\\
18 & -0.2913816  & -0.291382  & -0.6472304  &  -0.647264
   & -0.7890476  & -0.789086
\\
19 & -0.2976095  & -0.297609  & -0.6567835  &  -0.656834
   & -0.8002204  & -0.800253
\\
20 & -0.3036106  & -0.303610  & -0.6659643  &  -0.666010
   & -0.8109905  & -0.811021
\\
21 & -0.3094018  & -0.309402  & -0.6748057  &  -0.674847
   & -0.8213925  & -0.821424
\\
22 & -0.3149982  & -0.314999  & -0.6833365  &  -0.683371
   & -0.8314567  & -0.831488
\\
23 & -0.3204133  & -0.320413  & -0.6915817  &  -0.691614
   & -0.8412095  & -0.841236
\\
24 & -0.3256590  & -0.325659  & -0.6995634  &  -0.699594
   & -0.8506741  & -0.850702
\\
25 & -0.3307463  & -0.330746  & -0.7073012  &  -0.707325
   & -0.8598713  & -0.859898
\\
26 & -0.3356848  & -0.335684  & -0.7148123  &  -0.714840
   & -0.8688193  & -0.868843
\\
27 & -0.3404836  & -0.340483  & -0.7221124  &  -0.722128
   & -0.8775347  & -0.877560
\\
28 & -0.3451508  & -0.345150  & -0.7292154  &  -0.729240
   & -0.8860324  & -0.886055
\\
29 & -0.3496937  & -0.349693  & -0.7361338  &  -0.736144
   & -0.8943255  & -0.894348
\\
30 & -0.3541191  & -0.354118  & -0.7428791  &  -0.742900
   & -0.9024264  & -0.902449
\\
\end{tabular}
\end{ruledtabular}
\label{table1}
\end{table}
\begin{table}
\caption{Quantum defects $\mu'=\mu-15$ for the lowest states with $E>0$ and
$l=0$ where $n_r'=n_r-15$.  Results are listed for 50 and 100 MHz harmonic
traps calculated using quantum defect theory (QDT) and the self-consistent
method (SC).}
\begin{ruledtabular}
\begin{tabular}{ldddd}
$n_r^{\prime}$   & \multicolumn{2}{c}{50 MHz} & \multicolumn{2}{c}{100 MHz} \\
  &  \multicolumn{1}{c}{QDT} &  \multicolumn{1}{c}{SC}  &
  \multicolumn{1}{c}{QDT}  &  \multicolumn{1}{c}{SC}   \\
\hline
$0$    &  -0.4487958  &  -0.449380
       &  -0.5764557  &  -0.578201    \\

$1$    &  -0.5537186  &  -0.554150
       &  -0.6897921  &  -0.690958    \\

$2$    &  -0.6208043  &  -0.621154
       &  -0.7650603  &  -0.765956    \\

$3$    &  -0.6716889  &  -0.672001
       &  -0.8236159  &  -0.824344    \\

$4$    &  -0.7134089  &  -0.713686
       &  -0.8724945  &  -0.873108    \\

$5$    &  -0.7491700  &  -0.749404
       &  -0.9149564  &  -0.915486    \\

$6$    &  -0.7807137  &  -0.780921
       &  -0.9528010  &  -0.953266    \\

$7$    &  -0.8090982  &  -0.809287
       &  -0.9871383  &  -0.987552    \\

$8$    &  -0.8350137  &  -0.835187
       &  -1.018702   &   -1.01907    \\

$9$    &  -0.8589405  &  -0.859100
       &  -1.048008   &   -1.04835    \\

$10$   &  -0.8812266  &  -0.881372
       &  -1.075435   &   -1.07574    \\

$11$   &  -0.9021318  &  -0.902268
       &  -1.101266   &   -1.10155    \\

$12$   &  -0.9218555  &  -0.921982
       &  -1.125723   &   -1.12598    \\

$13$   &  -0.9405553  &  -0.940675
       &  -1.148980   &   -1.14922    \\

$14$   &  -0.9583577  &  -0.958469
       &  -1.171182   &   -1.17141    \\

$15$   &  -0.9753662  &  -0.975472
       &  -1.192445   &   -1.19266    \\

$16$   &  -0.9916661  &  -0.991766
       &  -1.212865   &   -1.21307    \\

$17$   &  -1.007329   &   -1.00742
       &  -1.232523   &   -1.23271    \\

$18$   &  -1.022416   &   -1.02251
       &  -1.251490   &   -1.25167    \\

$19$   &  -1.036879   &   -1.03706
       &  -1.269826   &   -1.26700    \\

$20$   &  -1.051062   &   -1.05114
       &  -1.287583   &   -1.28774    \\

$21$   &  -1.064704   &   -1.06478
       &  -1.304806   &   -1.30496    \\

$22$   &  -1.077940   &   -1.07801
       &  -1.321536   &   -1.32168    \\

$23$   &  -1.090799   &   -1.09087
       &  -1.337806   &   -1.33794    \\

$24$   &  -1.103309   &   -1.10338
       &  -1.353649   &   -1.35378    \\

$25$   &  -1.115492   &   -1.11556
       &  -1.369091   &   -1.36922    \\

$26$   &  -1.127370   &   -1.12743
       &  -1.384160   &   -1.38428    \\

$27$   &  -1.138963   &   -1.13902
       &  -1.398876   &   -1.39899    \\

$28$   &  -1.150287   &   -1.15035
       &  -1.413262   &   -1.41338    \\

$29$   &  -1.161358   &   -1.16142
       &  -1.427336   &   -1.42744    \\

$30$   &  -1.172191   &   -1.17225
       &  -1.441113   &   -1.44122    \\
\end{tabular}
\end{ruledtabular}

\label{table2}
\end{table}
\endgroup
\newpage

\begin{table}
\caption{Scaled energy eigenvalues $n^*$ for the lowest states with $E>0$ and
$l=2$ where $n_r'=n_r-14$.  Results are listed for a 10 MHz harmonic trap
calculated using quantum defect theory (QDT) and the generalised
self-consistent method (GSC).}
\begin{ruledtabular}
\begin{tabular}{lddldd}
$n_r'$   & \multicolumn{2}{c}{$n^{*}$} & $n_{r}$ &
\multicolumn{2}{c}{$n^{*}$} \\
  &  \multicolumn{1}{c}{QDT}  &  \multicolumn{1}{c}{GSC}   &  &
  \multicolumn{1}{c}{QDT}  &  \multicolumn{1}{c}{GSC}  \\
\hline

0 & 1.749908   &  1.749912  &   16  & 17.74052   &  17.74052  \\
1 & 2.749755   &  2.749759  &   17  & 18.73948   &  18.73949  \\
2 & 3.749538   &  3.749542  &   18  & 19.73840   &  19.73841  \\
3 & 4.749259   &  4.749263  &   19  & 20.73728   &  20.73728  \\
4 & 5.748920   &  5.748924  &   20  & 21.73610   &  21.73610  \\
5 & 6.748522   &  6.748526  &   21  & 22.73488   &  22.73489  \\
6 & 7.748065   &  7.748069  &   22  & 23.73362   &  23.73363  \\
7 & 8.747551   &  8.747555  &   23  & 24.73232   &  24.73232  \\
8 & 9.746982   &  9.746986  &   24  & 25.73097   &  25.73097  \\
9 & 10.74636   &  10.74636  &   25  & 26.72958   &  26.72959  \\
10 & 11.74568  &  11.74568  &   26  & 27.72816   &  27.72816  \\
11 & 12.74495  &  12.74495  &   27  & 28.72669   &  28.72669  \\
12 & 13.74416  &  13.74417  &   28  & 29.72518   &  29.72519  \\
13 & 14.74333  &  14.74330  &   29  & 30.72364   &  30.72365  \\
14 & 15.74244  &  15.74244  &   30  & 31.72207   &  31.72207  \\
15 & 16.74150  &  16.74151  & & &  \\

\end{tabular}
\end{ruledtabular}
\label{table3}
\end{table}


\end{document}